\documentclass[%
 reprint,
 amsmath,amssymb,
 aps,
]{revtex4-1}

\usepackage{color}

\usepackage{hyperref}
\usepackage{graphicx}
\usepackage{dcolumn}
\usepackage{bm}
\usepackage{color}

\begin{document}

\preprint{APS/123-QED}

\title{Brain Network Architecture: Implications for Human Learning}

\author{Marcelo G. Mattar$^{1,2}$}
\author{Danielle S. Bassett$^{1,3,4}$}
 \email{dsb@seas.upenn.edu}
\affiliation{
 $^1$Department of Bioengineering, University of Pennsylvania, Philadelphia, PA, 19104
}
\affiliation{
$^2$Department of Psychology, University of Pennsylvania, Philadelphia, PA, 19104
}
\affiliation{
 $^3$Department of Electrical and Systems Engineering, University of Pennsylvania, Philadelphia, PA, 19104
}
\affiliation{
$^4$To whom correspondence should be addressed: dsb@seas.upenn.edu
}

\date{\today}
\begin{abstract}
Human learning is a complex phenomenon that requires adaptive processes across a range of temporal and spacial scales. While our understanding of those processes at single scales has increased exponentially over the last few years, a mechanistic understanding of the entire phenomenon has remained elusive. We propose that progress has been stymied by the lack of a quantitative framework that can account for the full range of neurophysiological and behavioral dynamics both across scales in the systems and also across different types of learning. We posit that network neuroscience offers promise in meeting this challenge. Built on the mathematical fields of complex systems science and graph theory, network neuroscience embraces the interconnected and hierarchical nature of human learning, offering insights into the emergent properties of adaptability. In this review, we discuss the utility of network neuroscience as a tool to build a quantitative framework in which to study human learning, which seeks to explain the full chain of events in the brain from sensory input to motor output, being both biologically plausible and able to make predictions about how an intervention at a single level of the chain may cause alterations in another level of the chain.  We close by laying out important remaining challenges in network neuroscience in explicitly bridging spatial scales at which neurophysiological processes occur, and underscore the utility of such a quantitative framework for education and therapy.
\end{abstract}

\maketitle

\section*{A taxonomy of human learning}

In its simplest conceptualization, the nervous system can be understood as a mapping device that selects motor commands in response to stimuli present in the world \cite{sejnowski1994computational}. The goal of this mapping is to optimize output behavior to ensure survival of the organism. However, the complexity of the inputs makes the mapping far from trivial: the state of the world is always changing, and the number of possible states of the world is virtually infinite. It is precisely for that reason that evolution's solution to this problem does not involve an infinitely long stimulus-response lookup table. Instead, it provides animals with the capacity to adapt and optimize the input-output mappings in various temporal scales within a life time, allowing animals to thrive in a highly complex and ever-changing world. This adaptability -- either an acquisition or modification of behaviors in response to the environment -- is what we call learning \cite{schunk1996learning}. Given its importance in optimizing survival, learning is implemented in the brain in a variety of ways, and so it is useful to draw distinctions with regards to the mechanism of adaptation, to the content being learned, and to whether the individual is aware of this process or not. 

The process of learning always involves a learner, who undergoes adaptation in stimulus-response mappings, and the process may or may not involve an instructor. The learner can be considered the individual as a whole, or can be considered a subset of the nervous system such as a specific brain region or even a specific neuronal population. The instructor, when it exists (``instructed learning''), influences learning by providing qualitative or quantitative information about the learning process \cite{knudsen1994supervised}. If the information provided is the correct input-output mapping, learning is said to be ``supervised'' -- a term drawn from computer science \cite{knudsen1994supervised,ayodele2010types}. If the instruction is a feedback signal in the form of a binary fact (i.e., right/wrong) or continuous reward (i.e., from bad to good), the process is called ``reinforcement learning'' \cite{niv2009reinforcement}. Finally, if learning occurs without the presence of an instructor, the process is called ``unsupervised'', ``non-instructed'', or ``discovery learning'' \cite{barlow1989unsupervised}. The instructor -- like the learner -- can be another individual (i.e., a teacher or a guardian), or it can be a different part or signal in the brain (e.g., dopamine which signals reward prediction errors in the brain). The spectrum of supervised, to reinforcement, to unsupervised learning often coincides with a spectrum of acquisition times, with instructed learning (and in particular supervised learning) happening more quickly than non-instructed, unsupervised learning. 

Information about the state of the world arrives as inputs to the system, often through sensory organs. The goal of the sensory systems in the brain is to faithfully recover information about the local environment. However, because a large amount of sensory information is constantly bombarding the brain, humans are consciously aware of only a few of these inputs at any given time. For this reason, learning does not always happen as a conscious process (``explicit learning''), but can also occur outside of awareness (``implicit learning'') \cite{seger1994implicit}. And despite the massive amount of information received through our sensory organs, sensory information alone is rarely sufficient to allow a complete inference about the state of the world or the content of one's surroundings. Therefore, animals, and humans in particular, often rely on statistical relationships between sensory inputs over time to infer unseen features of the external world that are critical for behavior. In ``statistical learning'', the statistical relationships between sensory stimuli allows perception to infer or extrapolate information about the world \cite{aslin2012statistical,turk2005automaticity}.

In humans, a particular type of sensory information that carries significant representational content is language. Indeed, language is widely used as an instruction signal, allowing learning to take place even before any behavior is executed (through supervised learning) \cite{williams2002children}. This particular type of instructed learning, commonly observed in education and psychotherapy \cite{ruge2013functional}, contrasts with reinforcement learning, where the learner interacts with an environment and receives feedback signals in a process involving trial and error. Although education is a very prevalent form of learning for children and adolescents, reinforcement learning is still the primary means by which humans and other animals without language learn early in life \cite{cole2013rapid}. 

But learning through reinforcement signals can be a computationally daunting task: maximizing future reward may require a long sequence of behaviors before any type of feedback is received, leading to ambiguity on the link between the feedback and the specific behaviors executed. Indeed, because a very large number of actions are possible at every moment, the number of complete policies (sequences of actions towards a goal) grows exponentially with the number of actions, making the problem of learning by trial-and-error quickly intractable. One solution to this problem, widely used in machine learning but likely also by humans, is to combine actions together into subroutines \cite{wymbs2012differential,ramkumar2016chunking} -- for an action with positive valence like brushing one's teeth, one can combine the actions \textit{grab toothbrush}, \textit{put toothpaste}, \textit{bring into mouth}, and \textit{repeat the brushing movements several times}, into a single \textit{brush teeth} action), reducing the computational complexity of the reward maximization problem. This solution is referred to as ``hierarchical reinforcement learning'' \cite{botvinick2012hierarchical,botvinick2014model,frank2012mechanisms}.

\section*{In need of a quantitative framework}

As in any scientific field, separating the topic of study into smaller categories makes the problem easier to address. However, it is important that the individual categories of learning that we described above converge at some point into a general description: one that offers a fundamental understanding of behavior, the primary goal of psychology and neuroscience. A major challenge hampering progress towards such a description is the lack of a quantitative framework that can account for the full range of neurophysiological and behavioral data across different types of learning. Such a framework would be critical for understanding the manner in which sensory information translates into the diversity and flexibility of behavior observed in humans. In other words, a framework that bridges different types of learning must bridge different scales of neurophysiology. Moreover, such a framework -- which brides sensory information and action -- would enable the design of interventions to reshape behavior according to individual needs. These interventions could involve a combination of approaches ranging from reward signals, instruction signals, implicit or explicit training, and certainly a good deal of self-discovery on the part of the individual \cite{valiant2014must}. 

It is interesting to note that a \emph{quantitative framework} can become a \emph{quantitative theory} of human learning, if bolstered by appropriate physiological and physical intuitions and evidence. But to understand how a framework can become a theory, let us consider exactly what such a theory is \emph{not}. First, a quantitative theory is not only a quantitative description of data (of any sort), but instead is a model that can be used to predict behavior. Second, a quantitative theory must go beyond phenomenological models that can predict behavior in a narrow range of environmental conditions but with no biological realism, as this class of models cannot offer insights into true mechanisms. Third, such a theory should comprise more than a specific subset of phenomena, perhaps useful for prediction and biophysically realistic, but not covering the full range of processes between stimuli and behavior. Indeed, a quantitative theory of human learning must comprise the entire chain of events in the brain from sensory input to motor output, being both biologically plausible and able to make predictions about how an intervention at a single level of the chain may cause alterations in another level of the chain. 

There is an enormous appeal in obtaining such a theory. From a behavioral perspective, it would allow an understanding of how identical sensory stimuli may lead to vastly different behavior in a single individual, depending on its environmental conditions or phase of life. The same theory may, in addition, account for variability across individuals, taking into consideration the learning environments and conditions that they differently experienced. From a neuroscientific perspective, it could provide a greater understanding of the physiological processes that reshape information as it is transferred from one layer of the nervous system to another \cite{craver2005beyond}. From an educational perspective, it could elucidate more effective teaching approaches leading students to conform to social norms or learn educational material more effectively \cite{immordino2012implications}. And from a clinical perspective, it holds the promise to reshape the way that interventions are made so that behavior is implicitly and explicitly altered to improve one's day-to-day experiences \cite{johnson2013neuromodulation}. Put simply, such a theory could offer a principled empirical framework in which to guide efforts in education, cognitive and behavioral therapy, and clinical interventions for the betterment of society.

\section*{Existing theoretical and empirical frameworks}

Clearly, the process of learning, by which behavior is optimized in response to sensory information, is a complex one, being implemented by the nervous system in a variety of ways. If we isolate the organism in terms of its input-output mapping device -- the brain --, it becomes clear that this mapping is far from trivial, involving numerous intermediate brain regions situated conveniently between sensory input and motor output areas. Some of these intermediate regions are useful for quantifying value \cite{sugrue2004matching,knutson2005distributed}, others signal sensory predictions or prediction errors \cite{schultz2000neuronal}, and even others represent the abstract meaning of language \cite{chai2016functional}. For this reason, we argue, it is not enough to isolate a single component of the system when studying how it affects behavior. Instead, a holistic view of learning \textit{must} comprise interacting parts and dynamics in its description, and must also culminate in a description of how the individual parts reconfigure specific patterns of neural activity in motor areas which, in turn, drive behavior.

Despite such complexity, neuroscience research has, historically, focused heavily on (i) relationships between neural activity and sensory stimuli \cite{olshausen1996emergence}, or (ii) between neural activity and behavior \cite{rizzolatti1996premotor}. A better approach, instead, should comprise the full set of relationships between parts to deal with how information is propagated through the brain. And in the specific case of learning, one should focus on how these relationships \textit{change} over time, to enable changes in behavior. Understanding how such a complex process such as learning unfolds in time, therefore, is hardly feasible without a mathematical framework that can quantify the interactions of the individual parts and the orchestration of behavioral adaptability \cite{bassett2006adaptive}. 

Yet, because a description of the full set of processes between sensory stimulation and behavior in terms of molecular signaling or spike trains is hardly feasible, the study of the neurophysiology of learning has historically focused on subcomponents of the system. In effect, this corresponds to a reassignment of the input-output layers to lower levels such as, for instance, presynaptic and postsynaptic neurons. Indeed, this approach has been successful in describing how learning shapes the communication between individual neurons. Consider the field of spike-timing dependent plasticity, which relates changes in the efficiency of communication between neurons in terms of relative spike timings \cite{sjostrom2010spike}. If presynaptic spikes occur just before postsynaptic spikes more often than the reverse, synapses are strengthen (this process is referred to as \emph{Hebbian learning}). In contrast, if presynaptic spikes occur more often following postsynaptic spikes, than synapses are weakened (\emph{anti-Hebbian learning}). These rules are now ubiquitous in the study of the neurophysiology of learning as they effectively describe learning at the level of neurons. Still, to bridge this level of description to the level of behavior, one must implicitly assume that by combining a series of individual synaptic changes from sensory areas to motor areas, behavior is ultimately reshaped.

\begin{figure}[t]
\begin{center}
\centerline{\includegraphics[width=0.45\textwidth]{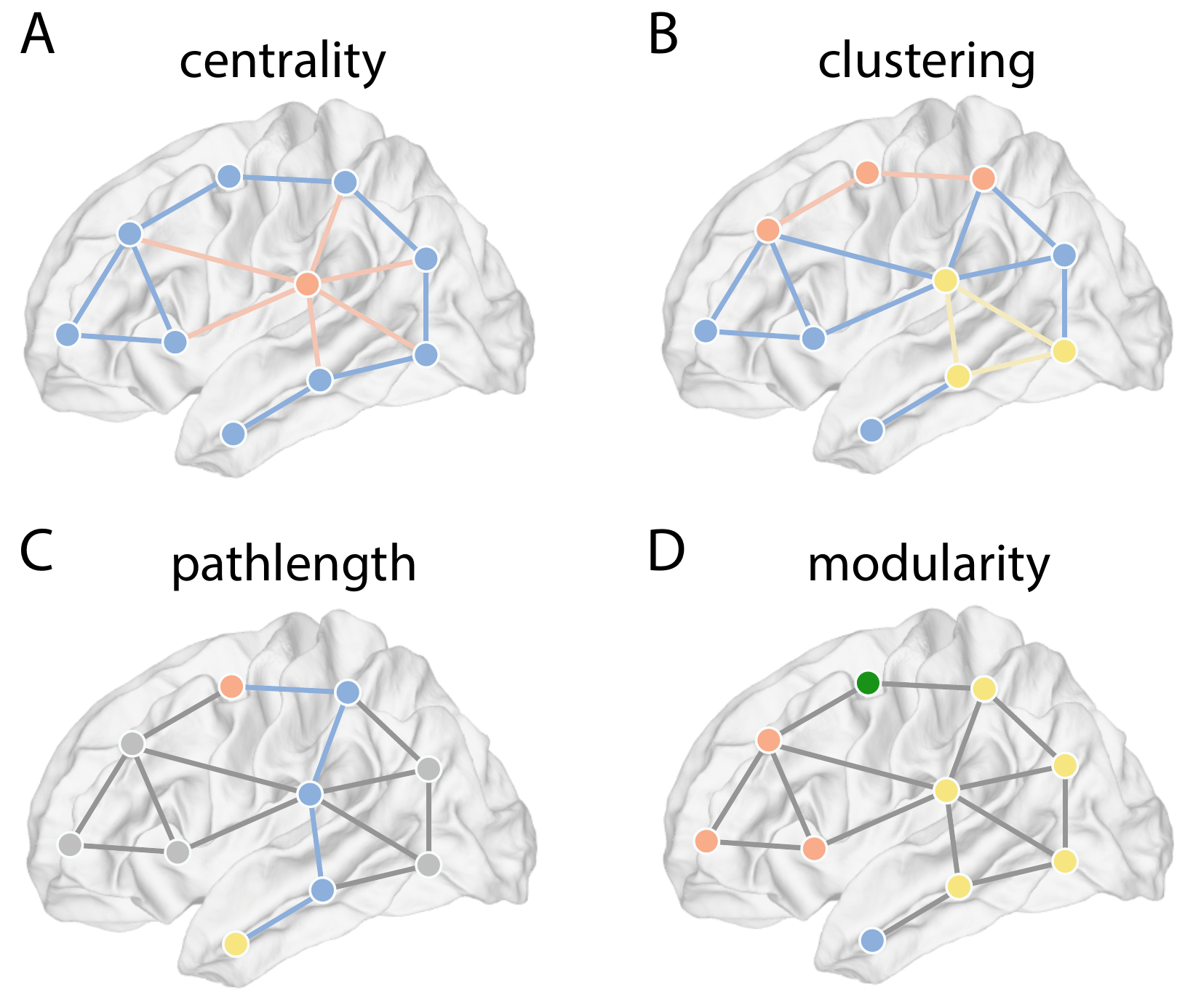}}
\caption{\textbf{Common statistical measures of network architecture.} Network architecture in complex systems can be studied by extracting summary statistics that each measure a different feature of that architecture. \emph{(A)} Network centrality is a notion of a node's influence on the network. Perhaps the most commonly studied centrality measure is \emph{degree centrality} which is simply the number (or strength) of edges emanating from a node. The node with the highest degree centrality in this graph is highlighted in peach, along with all of its edges. High degree nodes are often referred to as \emph{hubs} in the network.  \emph{(B)} The \emph{clustering coefficient} is a measure of local architecture in a network, and is commonly estimated by determining the ratio of the number of closed triangles (here highlighted in yellow) to connected triples (here highlighted in peach). The higher this ratio, the higher the clustering coefficient of the network, and the greater the expected potential for local processing. \emph{(C)} The shortest \emph{pathlength} between two nodes is defined by the fewest number of edges that must be traversed in order to get from the first node to the second node. The shortest path between the first node (peach) and the second node (yellow) is highlighted in blue; all other nodes and edges not participating in the shortest path are represented in grey.  Networks with relatively short pathlenths on average between all nodes in the network are thought to be efficient at transmitting information globally throughout the system. \emph{(D)} A network is said to display community or \emph{modular structure} if it contains groups of nodes that are more densely interconnected to one another than they are to other groups. In this graph, there exists a peach module (left) and a yellow module (right), in addition to two singletons (highlighted in blue and green, respectively).  Networks with community or modular structure are thought to effectively segregate processing within the separate modules. } \label{fig_net_stats}
\end{center}
\end{figure}

Another quantitative model of the neurophysiology of learning that considers feedback signals is reward-dependent plasticity \cite{beitel2003reward}. This model builds on spike-timing dependent plasticity, but proposes that the magnitude of synaptic change is modulated by the presence of reward signals -- commonly the neurotransmitter dopamine. If this neurotransmitter -- thought to signal a reward prediction error -- is present in the system, plasticity is relatively more likely to happen then when it is not present. Through reward-dependent plasticity, the nervous system can be reshaped so that reward prediction improves and prediction error decreases over time. Again, this mechanistic description of learning is effective in capturing essential aspects of neuronal plasticity, but falls short of describing the full set of interactions in the brain that occur with learning.

One can also examine the effects of learning at larger scales. Since the advent of neuroimaging, region-to-region interactions can be quantified as statistical dependencies between the blood-oxygen-level dependent (BOLD) signals from two regions, enabling a large-scale quantification of the effects of learning. These statistical dependencies are often referred to as estimates of so-called \emph{functional connectivity} \cite{friston2011functional}, because of their potential relationship to inter-regional communication \cite{fries2005mechanism,fries2015rhythms}. For example, in a study of explicit learning, functional connectivity was found to be higher between the medial-temporal lobe and dorsolateral prefrontal or lateral occipital cortices; in contrast, during implicit learning, functional connectivity is more pronounced between the medial temporal lobes and the thalamus \cite{mcintosh2003functional}. Further illustrating the potential for these approaches, it was observed that, in a reinforcement learning paradigm, the dorsal and ventral striatum are differentially functionally connected to the areas near the substantia nigra and ventral tegmental area, respectively \cite{kahnt2009dorsal}. These techniques, therefore, allowed the researcher to determine how a single brain area is interacting with a diverse set of other regions, allowing one to infer their potential role in the broader context of learning.

\section*{Complex systems science: a useful framework for multiscale neurophysiological processes}

Yet, to get the full picture of input-output mappings, bridging across types of learning and individual paradigms, one needs a holistic picture of brain functioning. Fortunately, a new field of science has emerged in the last decade to describe how parts of a system give rise to its collective behaviors \cite{newman2010networks}. The framework of complex systems deals with systems that display emergent behavior: behavior that cannot be explained by the dynamics of individual system components, but instead arises from a complex pattern of interactions between the parts.

To take a concrete example in the brain, while many sensorimotor, subcortical, and frontal regions change their activity as someone learns a new motor skill \cite{dayan2011neuroplasticity}, none of these changes -- by themselves -- could lead to learning.  Instead, changes in the processing performed by one brain region can alter the processing that is performed in other brain regions. As these alterations build on one another, a pattern of interactions across the brain can emerge, enabling a change in behavior \cite{friston2002beyond,happel2016dopaminergic}. While a comprehensible account of the collective changes in terms of individual interactions may appear daunting, one can turn to the mathematics of complex systems to obtain a qualitative and quantitative understanding of the system at various levels of description.

Complex systems science is a relatively broad discipline, and not all of its tools are necessarily relevant for understanding the brain. However, a particular subdiscipline known as network science does appear particularly relevant because it offers a mathematical framework to characterize systems composed of heterogeneously interacting parts (Fig.~\ref{fig_net_stats}). Built on the mathematical field of graph theory, network science has previously been used to describe properties of social networks \cite{granovetter1973strength,travers1969experimental}. Due to its general formulation, however, the majority of its techniques have been extended to describe systems as varied as smart-phone networks \cite{onnela2016harnessing,torous2016new}, granular materials \cite{bassett2015extraction,bassett2012influence,lopez2013jamming}, microbial communities \cite{steinway2015inference,baldassano2016topological}, or genetic expression \cite{zanudo2015cell,conaco2012functionalization}.

\begin{figure*}[t]
\begin{center}
\centerline{\includegraphics[width=0.70\textwidth]{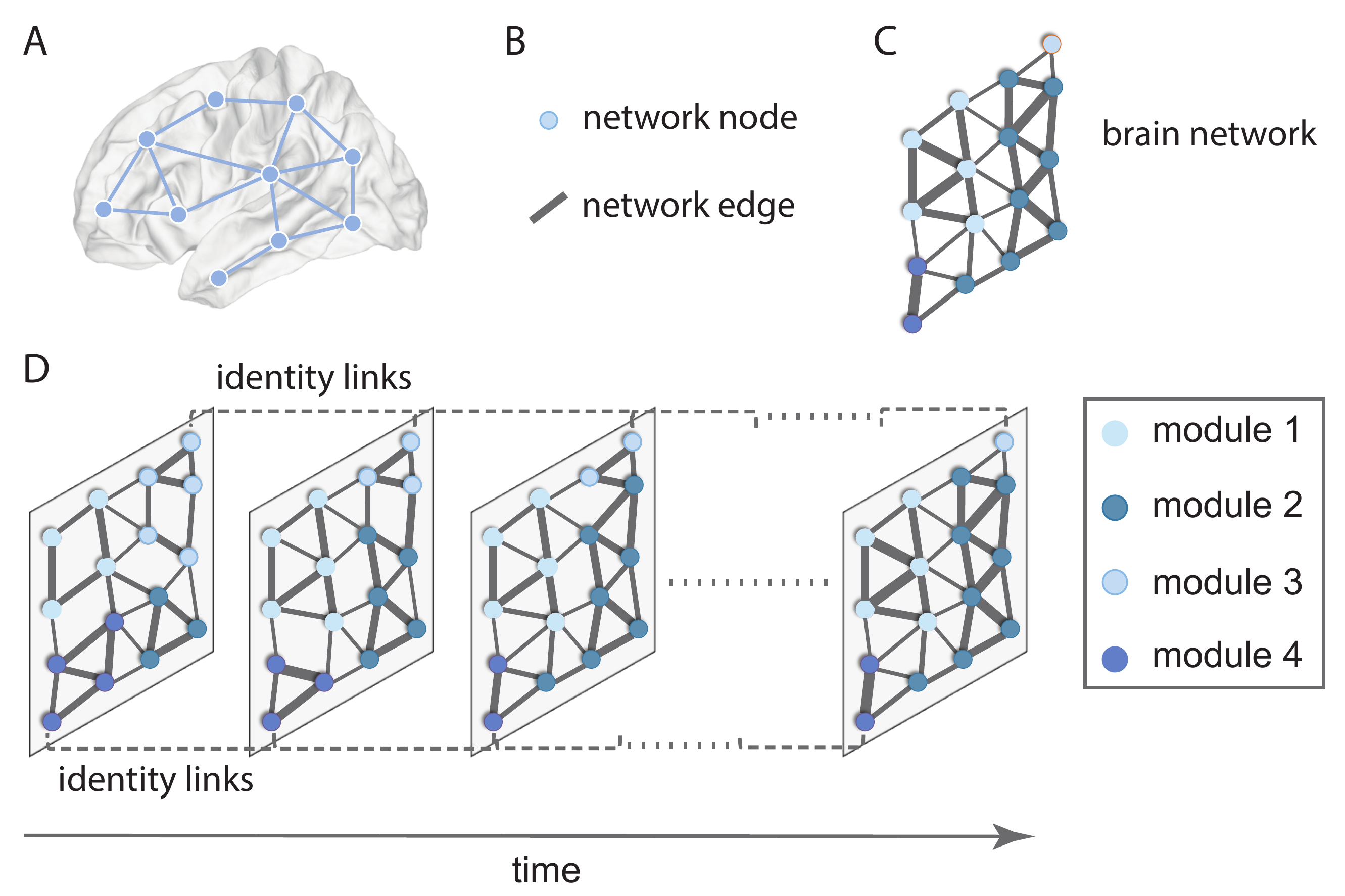}}
\caption{\textbf{Network and multilayer network representations of the brain.} \emph{(A)} A brain graph or network. \emph{(B)} The two fundamental units of the brain graph are nodes (brain regions) and edges (functional or structural relationships or links between nodes). \emph{(C)} Together, the pattern of edges interconnecting nodes forms a graph or network, from which we can extract meaningful patterns, principles, and predictions. \emph{(D)} In many adaptive processes, complex networked systems like the brain display patterns of interconnectivity that change over time. Particularly in learning, functional brain networks reconfigure as behavior changes. To address this temporal complexity, multilayer networks represent a brain network estimated over a particular time window as a single layer inside of a larger system. Networks in one layer are explicitly linked to networks in the preceding layer and the immediate future layer by so-called \emph{identity links}. These identity links enable the network to be studied as a single entity, rather than an ensemble, significantly simplifying the mathematics and remaining true to the inherent dependencies between time windows (layers). Multilayer networks can be used to represent time-varying connectivity patterns in the brain as someone learns, and such applications have uncovered non-trivial reconfigurations of functional modules supporting learning. } \label{fig_multilayer}
\end{center}
\end{figure*}

The application of network science to study the brain typically entails a parcellation of the brain into a set of regions that display roughly independent structure or function (Fig.~\ref{fig_multilayer}A). The relationship between these regions can then be characterized based on structure (i.e. physical connections) or function (i.e. statistical relationships between regional timecourses) (Fig.~\ref{fig_multilayer}B). Mathematically, regions of the brain are defined as the nodes of a graph, and their relationship as edges connecting the corresponding nodes (Fig.~\ref{fig_multilayer}C). A graph with $n$ nodes can be represented as an $n \times n$ adjacency matrix $A_{ij}$, whose element $a_{ij}$ corresponds to the connection strength between node $i$ and node $j$. Once the adjacency matrix of a graph is defined, one can then draw on techniques from graph theory and network science to extract emerging properties of the system \cite{bollobas1985random,bollobas1979graph}. 

One of the main advantages of the network science approach in neuroscience lies in its ability to detect structures at the mesoscale level -- encompassing subsets of network nodes \cite{bassett2006small,bassett2016small}. One of the most important emerging network properties at the mesoscale level is community or modular structure \cite{girvan2002community}. A community (or module) is defined as a subset of nodes from the graph that are more tightly connected to each other than they are to the rest of the network. Many real networks exhibit some degree of community structure. In the brain, interestingly, these clusters display beautiful overlap with known cognitive systems, including motor, visual, auditory, default mode, salience, attention, executive, and subcortical systems \cite{chen2008revealing,meunier2009age,power2011functional,yeo2011organization,gordon2016generation}. Importantly, these modules are differentially expressed, and show different patterns of interaction with one another while a human is learning \cite{bassett2015learning}. 

Identifying the community structure of a network is a complex problem, though, and most of the current approaches rely on heuristic algorithms to address an NP-complete problem. A particularly successful technique is the optimization of the modularity quality function, which involves partitioning network nodes into modules (or communities) such that the total connection strength within groups of the partition is more than would be expected in random networks of similar kind \cite{newman2006modularity}. The complexity of the approach resides in the fact that each partition needs to be tested in a brute-force manner.

While this approach may at first seem too abstract and simplistic, it captures exactly what have been listed as desirable features of a quantitative framework of learning: a holistic description of region-to-region interactions encompassing the whole brain, applicable across tasks and experimental paradigms, allowing one to identify emerging properties and features in a quantitative manner. Indeed, it is precisely the simplicity and generality of this approach that makes it so powerful, and its usefulness is evident from the vast number of insights brought by this burgeoning field now termed "Network Neuroscience" \cite{muldoon2014network,medaglia2015cognitive,misic2016from,bassett2016network}.

\section*{Network neuroscience and its implications for human learning}

A single network can assume different forms over time or across different conditions. For example, a network of friends can evolve over time, a network of blogs can be different in periods of election or in times of major sports events, and a network of brain regions can assume different forms before and after the acquisition of a new skill. In all cases, the interpretation of each condition or time point as a separate network obscures the fact that some nodes or edges have the same identity across networks. Multilayer networks can account for this emerging structure by assuming that nodes or edges can change across conditions or time, while still maintaining their identity \cite{kivela2014multilayer}. In these cases, a multilayer network with $n$ nodes and $p$ layers can be represented as an $n \times n \times p$ tensor $A_{ijs}$, where element $a_{ijs}$ corresponds to the connection strength between node $i$ and node $j$ in layer (or slice) $s$ \cite{holme2012temporal}. 

The application of network neuroscience to the study of human learning, in particular, benefits greatly from the mathematical formulation of multilayer networks \cite{bassett2011dynamic} (Fig.~\ref{fig_multilayer}D). For a given set of nodes, if their relationships -- network edges -- are measured at different points in time while learning unfolds, the corresponding multilayer network can capture the reconfigurations that the entire brain undergoes. This approach has been widely successful, in particular, in the case of functional networks, which are sensitive enough to variations on the timescale of a single experiment. These graphs can be extracted during task performance on a single day \cite{telesford2016detection,braun2015dynamic,mattar2015functional} or across many days in longitudinal study designs \cite{bassett2011dynamic,bassett2015learning}.

A particularly interesting development in network neuroscience was the observation that several network metrics seem to change with learning \cite{bassett2011dynamic}. In particular in the context of motor skill learning, the connection strengths seem to increase as well as the average clustering coefficient, along with reduced communication distances and altered network centrality \cite{heitger2012motor,mantzaris2013dynamic} (for a careful description of these statistics, see \cite{rubinov2010complex}). The joint increase of connection strengths and clustering seem to suggest that the network architecture becomes more modular with learning -- a proposition that has received recent empirical support \cite{sporns2016modular}.

\begin{figure}[h]
\begin{center}
\centerline{\includegraphics[width=0.50\textwidth]{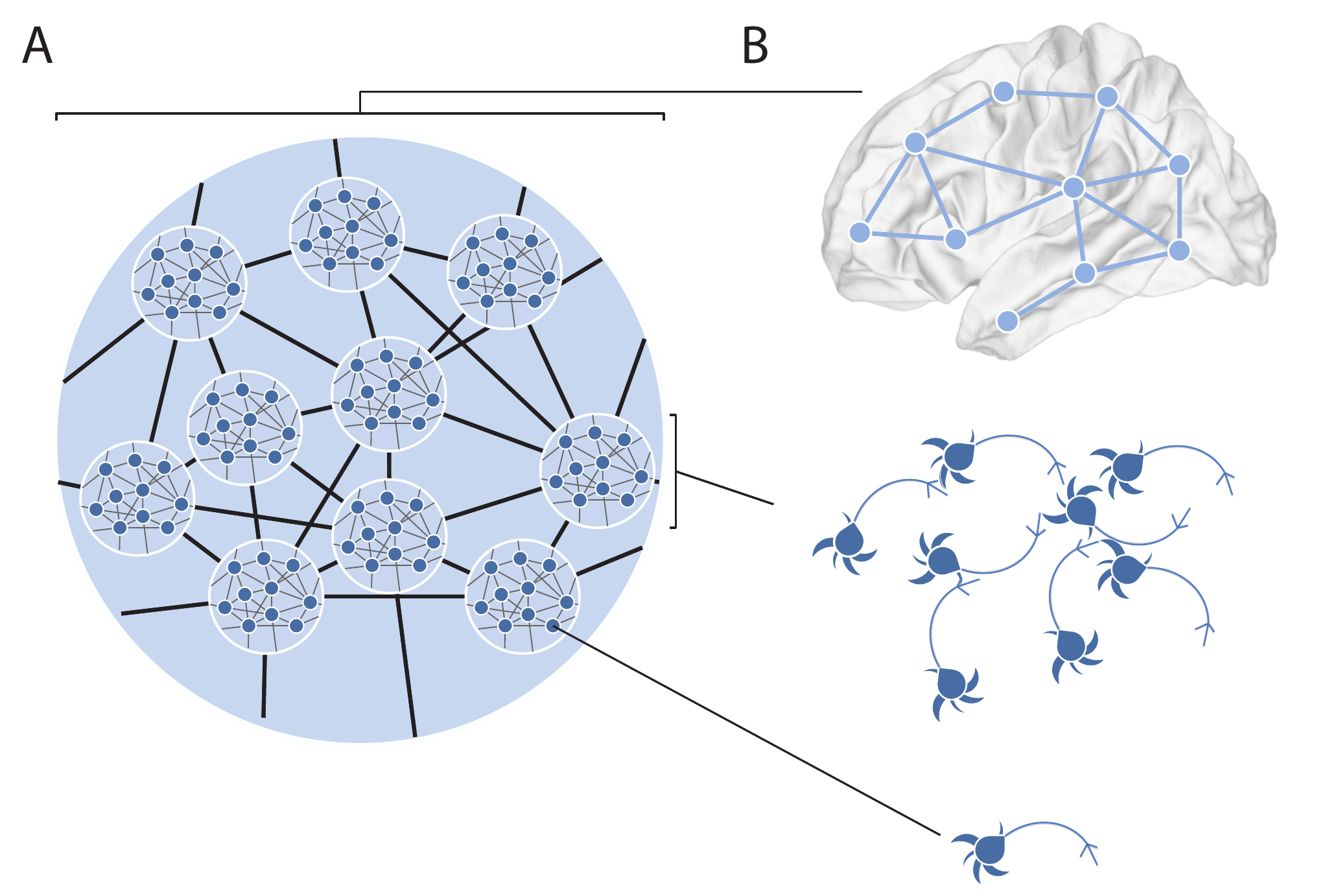}}
\caption{\textbf{The brain exhibits a hierarchical structure, where smaller units combine in groups and then into increasingly larger units.} \emph{(A)} In the network formulation, nodes and edges at the lowest levels combine into cohesive units that achieve some form of computation. These units, in turn, can be similarly thought of as nodes, with corresponding edges representing interactions with similarly sized units that similarly perform some computation. This hierarchical structure can be conceived at many distinct levels according to the computations considered. \emph{(B)} Each level of the hierarchy can also be thought of in biological terms. At the lowest level, nodes can represent individual neurons and edges the corresponding synapses. At higher levels, nodes can represent circuits or even entire brain regions that perform a meaningful form of computation.} \label{fig_multiscale}
\end{center}
\end{figure}

But how can one detect time-varying modular structure in multilayer networks? The basic approach consists of deriving an expanded single-layer network derived from the multilayer network by connecting each node to itself across layers. In time-evolving networks, this is done by connecting each node to itself across layers corresponding to adjacent time-points. In multislice networks, this is achieved by connecting each node to itself in all layers corresponding to different events or conditions \cite{mucha2010community}. Communities in multilayer networks can then be detected by optimizing the modularity quality function with the expanded $(np \times np)$ adjacency matrix. While a computationally challenging task \cite{bassett2013robust}, identifying communities in multilayer networks derived from neuroimaging data has proven particularly useful in the study of motor skill learning, where motor and visual modules become increasingly autonomous from one another with practice, as regions known to affect cognitive control release from the remainder of the network \cite{bassett2015learning}.

Once the community structure of a network is determined, several other network metrics can be defined \cite{chai2016functional}. For instance, insights into the role of individual nodes of the network can be obtained by calculating the fraction of its connections (or the fraction of the edge weights) that connect it to nodes within or between communities \cite{guimera2005functional}. For multilayer networks specifically, the module allegiance of a network summarizes the co-occurrence of network nodes in communities across layers such as individuals or experimental conditions \cite{bassett2015learning,mattar2015functional}. Similarly, the flexibility of a given brain region can be quantified as the number of times it changes community affiliation. Interestingly, individual differences in this flexibility of modular architecture predicts future learning rate in a motor-skill learning task \cite{bassett2011dynamic}. 

Overall, there is increasing evidence that modularity is intimately linked to learning. While the specific areas that display increased modularity with learning change from task to task, the increased organization of the network into distinct and roughly independent compartments seems to be a ubiquitous finding. While a relatively new finding in the context of brain network architecture and human learning, the relationship between modularity and adaptability is one that has been touted as fairly fundmental in many other disciplines \cite{kirschner1998evolvability,kashtan2005spontaneous}. In the context of the brain, modularity -- and particularly flexible changes in brain network organization -- may enable learning by enhancing memory \cite{stevens2012functional,braun2015dynamic} and other dimensions of executive function including cognitive flexibility \cite{braun2015dynamic}.

\section*{A quantitative framework will engage multiscale network neuroscience}

At this point, it is worthwhile to look back at the desirable features of a quantitative theory for learning and re-evaluate where we stand in its achievement. As stated previously, it should comprise a quantitative description of the interactions among system components, encompassing the entire chain of events in the brain from sensory input to motor output -- that is, behavior \cite{bassett2016TICSnetwork}. Moreover, it should detect emergent properties of the brain at multiple hierarchical levels and account for the variety of results obtained from different types of learning with a single framework. Taken together, it is clear that the framework of network neuroscience seems, indeed, to be a great candidate for achieving such a theory. Yet, being a candidate to achieve a theory is not the same as explicitly being the theory itself. 

Indeed, considering that we are still not able to design effective educational and therapeutic interventions based on the network neuroscience of learning, we must assume that we do not have a quantitative theory of human learning. It is worthwhile, therefore, to take a step back and evaluate our current approaches in light of our current understanding and ask if we have at least attained an appropriate quantitative \emph{framework}. An appropriate framework is one that can bridge levels of description from neural processes to human behavior. Can network neuroscience -- as an approach -- be this bridge \cite{bassett2016network}? Some argue that the approach in its current instantiation is incomplete or insufficient for a neurobiological description of human behavior. Indeed, critics point to the fact that network nodes are often ill-defined, grouping regions with a variety of structural and functional properties into a single chunk. Similarly, collapsing the diversity of interactions as an edge within a single dimension of variability may, at first, appear too simplistic. Are there ways to extend network neuroscience to meet these challenges?

Recent mathematical advances, methodological developments, and empirical evidence suggests that this is indeed the case. Network neuroscience in principle can be extended to any spatial scale of relational data, but even more importantly, can be used to simultaneously model multiple scales, and indeed to link the network representation in one scale to that in another. While these tools still simplify the full richness of the brain's structure and function into a graphical framework, they nevertheless directly address the repeated, hierarchical nature of the brain, thereby potentially providing the desired holistic description. And perhaps even these simplifications are important. From a philosophical perspective, to bridge between levels of description, one needs to abstract away the details of each particular level and identify the commonalities between them -- requiring an abstract notion of units (e.g. nodes) and their interactions (e.g. edges) (Fig.~\ref{fig_multiscale}).

Bridging between levels of description, we believe, is precisely what is missing in the current frameworks \cite{bassett2016network}. Just like a whole-brain description at the regional level allows the observation of emerging structure at the mesoscale level, we can infer that descriptions at different levels may allow the detection of emerging structures of different kinds. Consider the field of neurophysiology, for example, widely successful in describing the biophysics of neurons and their plasticity with learning. Bridging the gap between neuronal-level descriptions and larger-scale, region-to-region interactions, would bring decades of research in neurophysiology in contact with recent discoveries in network neuroscience \cite{betzel2016multiscale}, allowing a connection between sensory input and behavioral output at various levels of description. And, we propose, the language of network neuroscience may be the perfect language to bridge this gap.

\section*{Acknowledgements}
We thank Ari Kahn for comments on an earlier version of this manuscript. The work was supported by the NSF BCS-1430087.  We also acknowledge support from the John D. and Catherine T. MacArthur Foundation, the Alfred P. Sloan Foundation, the Army Research Laboratory and the Army Research Office through contract numbers W911NF-10-2-0022 and W911NF-14-1-0679, the National Institute of Health (2-R01-DC-009209-11, 1R01HD086888-01, R01-MH107235, R01-MH107703, and R21-M MH-106799), the Office of Naval Research, and the National Science Foundation (BCS-1441502, CAREER PHY-1554488, and BCS-1631550). The content is solely the responsibility of the authors and does not necessarily represent the official views of any of the funding agencies.

\bibliography{bibfile}

\end{document}